\documentstyle[11pt,epsf]{article}
\begin{document}

\par
\topmargin=-1cm      % distance from top of the page to first line
                     % of text plus one inch
%\vspace*{.5in}

{ \small
\noindent{U.Md. PP\# 99-030}\hfill{DOE/ER/40762-162}}

\vspace{40.0pt}

\begin{center}
{\large {\bf Scheming in Dimensional Regularization}}\\

\vspace{18pt}
{\bf D.R.~Phillips\footnote{Present address:
Department of Physics, University of Washington,
Box 351560, Seattle, WA, 98195-1560, Email: phillips@phys.washington.edu}
and S.R.~Beane\footnote{Email: sbeane@physics.umd.edu}}

\vspace{6pt}
Department of Physics, University of Maryland, College Park, MD, 
20742-4111

\vspace{12pt}
{\bf M.C.~Birse \footnote{Email: mike.birse@man.ac.uk}}

\vspace{6pt}
Theoretical Physics Group, Department of Physics and Astronomy\\ 
University of Manchester, Manchester, M139PL, U.K.

\end{center}
\vspace{12pt}

\begin{abstract}

We consider the most general loop integral that appears in
non-relativistic effective field theories with no light particles.
The divergences of this integral are in correspondence with simple
poles in the space of complex space-time dimensions. Integrals
related to the original integral by subtraction of one or more poles
in dimensions other than $D=4$ lead to nonminimal subtraction
schemes. Subtraction of all poles in correspondence with ultraviolet
divergences of the loop integral leads naturally to a regularization
scheme which is precisely equivalent to cutoff regularization. We
therefore recover cutoff regularization from dimensional
regularization with a nonminimal subtraction scheme. We then discuss
the power-counting for non-relativistic effective field theories
which arises in these alternative schemes.
\end{abstract}

\newpage

\section{Divergent integrals in non-relativistic effective field theories}

Effective field theories of non-relativistic scattering have been a
subject of much interest recently. These field theories have
applications in many branches of physics.  They are useful
in analyzing a system with widely separated energy scales. For instance, to
describe two-body scattering at momenta considerably lower than the
mass of any exchanged particle an effective field theory can be
written in which all of the exchanged particles are ``integrated out''
of the original field theory Lagrangian. In the low-energy effective
theory the effects of these ``heavy'' degrees of freedom are
represented by an expansion of the interaction Lagrangian as a
sequence of local operators of increasing dimension.  This low-energy
effective theory can then be used to calculate non-relativistic
scattering in the system of interest.

In pursuing such a calculation one encounters divergences.
These divergences must be regulated and renormalized before physical
quantities can be calculated. Here we consider the most general loop
integral appearing in a low-energy EFT description of non-relativistic
scattering. We regularize this integral using the standard technique
of dimensional regularization (DR). Within the framework of DR, we
identify the poles as a function of the number of space-time
dimensions that correspond to the divergences of this integral.  This
allows us to clarify the relation between DR-based schemes and cut-off
regularization. We also demonstrate explicitly how the power-counting
for the coefficients appearing in the non-relativistic effective
theory Lagrangian is affected by the choice of regulator.

Consider a non-relativistic effective field theory for the interaction
of two identical heavy particles of mass $M$. If no exchanged degrees
of freedom appear explicitly in the Lagrangian then we can immediately
write:

\begin{equation}
{\cal L}=\psi^\dagger i \partial_t \psi + \psi^\dagger \frac{\nabla^2}{2 M} 
\psi - \frac{1}{2} C_0 (\psi^\dagger \psi)^2\\
-\frac{1}{2} C_2 (\psi^\dagger \nabla^2 \psi) (\psi^\dagger \psi) + h.c. 
+ \ldots.
\label{eq:lag}
\end{equation}
where the dots refer to other invariants with two or more derivatives.
Of course, without a power-counting scheme this Lagrangian is useless,
since there are infinitely many interaction terms consistent with
assumed symmetries. A power-counting scheme must be established which
allows us to keep only a finite number of operators at a given order
in the gradient expansion. Unless this is done the effective field
theory will have no predictive power, as an infinite number of
coefficients will enter the calculation.

We can formally write the all-orders solution of the low-energy
effective theory by constructing the non-relativistic potential

\begin{equation}
V(\hat{p})=\sum_{n=0}^{\infty} C_{2n} \hat{p}^{2n}
\label{eq:potential}
\end{equation}
directly from the Lagrangian. Here $\hat{p}$ is to be understood as an
operator, which may denote momentum or energy-dependence. More
explicitly, this potential is

\begin{equation}
V({\bf k}',{\bf k};E)=\sum_{n=0}^\infty \sum_{i=1}^n \sum_j
C_{2n}^{(i,j)} p^{2(n-i)} {\cal O}^i_j({\bf k}',{\bf k}),
\end{equation}
where $p^2=ME$ and the index $j$ enumerates members of the set of all
operators of dimension $2i$ which are consistent with Hermiticity and 
rotational invariance. In the case of $s$-wave scattering, the operators 
that contribute are

\begin{equation}
{\cal O}_j^i=k^{2j}k'^{2(i-j)}+k^{2(i-j)}k'^{2j},
\end{equation}
although, as we shall see, these details of the construction of the
potential are not necessary for our discussions here. Note that the
potential $V$ contains only terms that are analytic in the momenta
$k$, $k'$ and $p$. This is because in writing the effective Lagrangian
(\ref{eq:lag}) we have made local expansions of any non-analytic
structures which appear in the ``full'' theory. Therefore, the only
non-analytic effects in this low-energy effective theory come from
solving the Schr\"odinger equation. The effective theory is therefore
only valid for energies well below the masses or production thresholds
of any exchanged particles.

\begin{figure}[h,t,b]
\vspace{0.5cm} 
\epsfysize=1.5cm 
\centerline{\epsffile{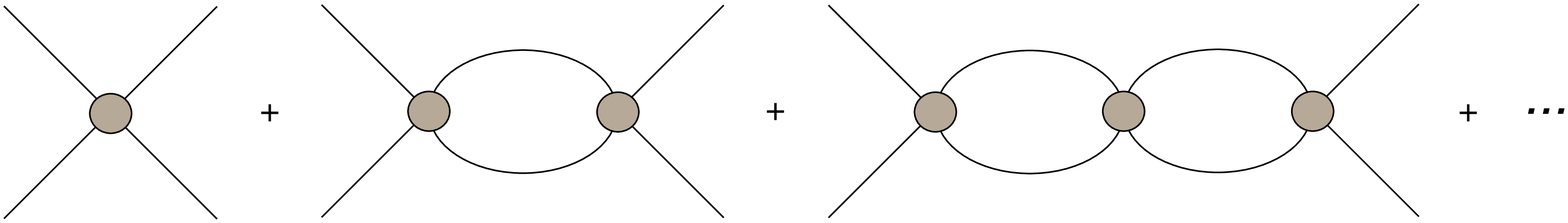}}
\centerline{\parbox{11cm}{\caption{\label{fig1} The diagrammatic
solution of the Lippmann-Schwinger equation with the effective
potential represented by the shaded blob. }}}
\end{figure}

The potential $V$ is then iterated via the Lippmann-Schwinger equation

\begin{equation}
T(k',k;E)=V(k',k;E) + M\int \frac{d^3q}{(2 \pi)^3} \, V(k',q;E) 
\frac{1}{EM- {q^2}+i\epsilon} T(q,k;E)
\label{eq:LSE}
\end{equation}
to give the on-shell scattering amplitude $T(p,p;E)$, with
$p=\sqrt{ME}$. All observables in the $\psi\psi$ system can be
obtained from this amplitude. In the language of Feynman diagrams, the
Lippmann-Schwinger equation generates the sum of all graphs, as
illustrated in Fig.~\ref{fig1}.

It immediately follows from Eq.~(\ref{eq:LSE}) that divergent
integrals will accompany every loop computed in the effective field
theory. This means that the $C_{2n}$ coefficients must be
renormalized, and therefore are renormalization scheme dependent.
Since the scaling behavior of the $C_{2n}$ coefficients determines
which operators are big and which are small in the effective theory,
this means that the power counting scheme will look different in
different regularization schemes.

In what follows we will consider the most general divergent integral
that appears in the effective theory.  In Section 2 we will explore
the many possible definitions of the divergent integral in dimensional
regularization, and then in Section 3 discuss the power counting for
the coefficients $C_{2n}$ that arise from these definitions.

The most general form of the divergent integral which appears 
in the effective theory is:

\begin{equation}
I_n(\mu)=\left(\frac{\mu}{2}\right)^{4-D} M \int \frac{d^{(D-1)}q}
{(2 \pi)^{D-1}} \frac{q^{2 n}}{ME^+ - q^2}
\label{eq:integral}
\end{equation}
where $D$ is the number of space-time dimensions, $\mu$ is a
renormalization scale, and $E^+=E + i \eta$, with $\eta$ a positive
infinitesimal, is the energy. Intuition about this integral
may be gained by evaluating with a sharp cutoff $\beta$. In $D=4$ this
gives

\begin{equation}
I_n(\beta)= M \int \frac{d^{3}q}
{(2 \pi)^{3}} \frac{q^{2 n}}{ME^+ - q^2}=
\frac{M}{2{\pi^2}}\int_0^\beta {dq} \frac{q^{2n+2}}{ME^+ - q^2}.
\label{eq:integralcutoffdef}
\end{equation}
The imaginary part is, of course, regularization scheme invariant.
Meanwhile, the real, or principal value, part is given by

\begin{equation}
{\rm Re} \, \, I_n(\beta)=
-\frac{M}{2{\pi^2}}\left( p^{2n} \beta + \frac{\beta^3}{3}p^{2n-2} +\ldots +
\frac{\beta^{2n+1}}{2n+1} \right)+
{p^{2n}}\frac{Mp}{4{\pi^2}}
\log\left(\frac{\beta +p}{\beta-p}\right),
\label{eq:integralcutoff}
\end{equation}
provided that $\beta^2 > p^2$. It is clear that the integral of
interest is ultraviolet power-law divergent at $D=4$. The maximum power
of the cutoff which appears is just the superficial degree of divergence
of the graph, $2n + 1$.

\begin{figure}[h,t,b]
\vspace{0.5cm} 
\epsfysize=5cm 
\centerline{\epsffile{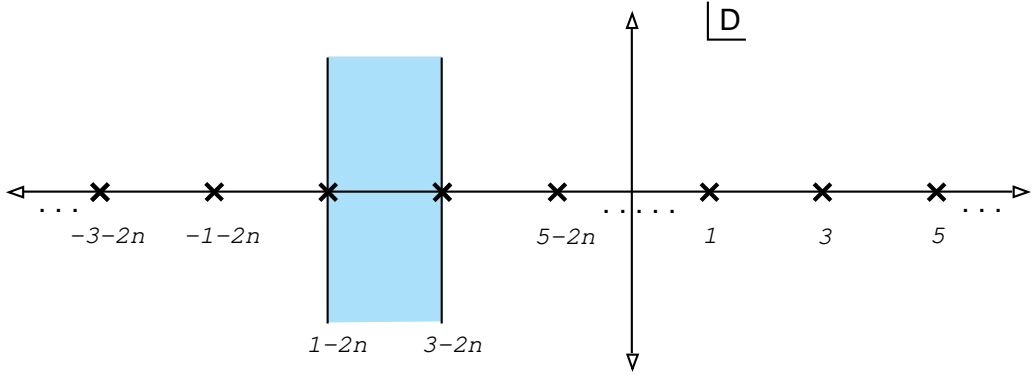}}
\centerline{\parbox{11cm}{\caption{\label{fig2} 
The complex $D$-plane.  The integral $I_n (\mu )$ is given by 
Eq.~(\protect{\ref{eq:result2}}) within
the shaded region of the $D$-plane.  Poles are represented by 
crosses.}}}
\end{figure}

Using the usual techniques employed in calculating dimensionally
regularized integrals in an arbitrary number of dimensions (see, for
instance, Appendix B of Ramond~\cite{Ra81}), the integral
(\ref{eq:integral}) may be re-expressed as,

\begin{equation}
I_n(\mu)=-\left(\frac{\mu}{2}\right)^{4-D} \frac{M}{(4 \pi)^{(D-1)/2}}
\frac{\Gamma(\frac{2n + D - 1}{2})}{\Gamma(\frac{D - 1}{2})}
(-M E)^{(2n + D - 3)/2} \Gamma\left(\frac{3 - D  - 2n}{2}\right)
\label{eq:result1}
\end{equation}
provided that $E < 0$. This equality is valid in a region of the
complex $D$-plane defined by

\begin{equation}
1 - 2n < D < 3 - 2n \mbox{ and } D > 1.
\label{eq:conditions}
\end{equation}
The first inequality is easy to understand from
Eq.~(\ref{eq:integral}) using simple power counting arguments.  The
condition $D > 1 - 2n$ ensures that integral is free of infrared
divergences and the condition $D < 3 - 2n$ ensures that the integral
is free of ultraviolet divergences.  According to
Eq.~(\ref{eq:result1}), in the complex $D$-plane (see Fig.~\ref{fig2})
these divergences manifest themselves as singularities of the gamma
functions in the corresponding regions.  The condition $D > 1$ follows
from requiring that the space-time measure be infrared safe.

The two conditions of Eq.~(\ref{eq:conditions}) are incompatible for
$n > 0$. Thus, in fact, to properly calculate the integral
(\ref{eq:integral}) we must do the angular integration in $D=4$, and
then consider the result of varying the number of dimensions in which
the ``radial'' integration is performed. This technique leads to the
result:

\begin{equation}
I_n(\mu)=-\left(\frac{\mu}{2}\right)^{4-D} \frac{M}{(4 \pi)^{3/2}}
\frac{\Gamma(\frac{2n + D - 1}{2})}{\Gamma(\frac{3}{2})}
(-M E)^{(2n + D - 3)/2} \Gamma\left(\frac{3 - D  - 2n}{2}\right),
\label{eq:result2}
\end{equation}
which is valid for 
\begin{equation}
1 - 2n < D < 3 - 2n,
\end{equation}
and $E < 0$.  If we are interested in scattering problems in four
space-time dimensions we must now perform two analytic continuations:
one in the energy $E$ and one in the number of dimensions
$D$. Performing the analytic continuation in $E$ first, by using the
evaluation at $E + i \eta$, rather than $E$ itself, we find that for
$E > 0$

\begin{equation}
I_n (\mu)=-\frac{M}{4 \pi^2} \left(\frac{\mu}{2}\right)^{4-D} (-ME)^n
(-ME^+)^{(D-3)/2} \Gamma\left(\frac{2n + D - 1}{2}\right) \Gamma
\left(\frac{3 - D -2n}{2}\right),
\label{eq:In}
\end{equation}
provided that $1 - 2n < D < 3 - 2n$. The following discussion 
makes extensive use of this result.

The usual dimensional regularization prescription for calculation of
the integral $I_n$ in dimensions where the expression (\ref{eq:In}) is
not valid is to define a new function $\tilde{I}_n (\mu)$
which agrees with the original $I_n(\mu)$ in the region $1 - 2n < D <
3 - 2n$, but is {\em defined by Eq.~(\ref{eq:In})} in the rest of the
complex $D$-plane. For {\em this} $\tilde{I}_n$ we may use the Gamma
function identity,

\begin{equation}
\Gamma(n+a) \Gamma(1-a-n)=\Gamma(a) \Gamma(1-a) (-1)^n,
\label{eq:identity}
\end{equation}
to obtain

\begin{equation}
\tilde{I}_n (\mu)=-\frac{M}{4 \pi^2} 
\left(\frac{\mu}{2}\right)^{4-D}
(ME)^n (-ME^+)^{(D-3)/2} \Gamma\left(\frac{D - 1}{2}\right)
\Gamma\left(\frac{3 - D}{2}\right),
\label{eq:InKSW}
\end{equation}
which is correct for all $D$ except odd integers.  This expression
differs from that of Ref.~\cite{Ka98A,Ka98B} by a factor due to the
way the angular integration was performed. Note that if the identity
(\ref{eq:identity}) is  applied directly to
Eq.~(\ref{eq:result1}), then the result of Refs.~\cite{Ka98A,Ka98B}
for $\tilde{I}_n(\mu)$ is recovered.

\section{Subtraction schemes and the divergent integral}

In $D=4$ Eq.~(\ref{eq:InKSW}) yields the result of Ref.~\cite{Ka96}:

\begin{equation}
\tilde{I}_n(\mu)=-p^{2n} \frac{i M p}{4 \pi},
\label{eq:mins}
\end{equation}
where $p=\sqrt{ME}$. The regularization scheme corresponding to
Eq.~(\ref{eq:InKSW}) with removal of any poles at the critical
dimension is, by convention, minimal subtraction. Of course, there are
no poles in the dimension of interest to us, $D=4$, and so only the
imaginary part of the integral survives in minimal subtraction.

The expression (\ref{eq:In}) has poles at $D=1-2n,-1-2n,\ldots$,
$D=3-2n,5-2n,7-2n, \ldots$. The former lie below the region of
analyticity of $\tilde{I}_n$ (see Fig.~\ref{fig2}), and can be considered
infrared divergences that arise when $I_n$ is evaluated in such
dimensions. Conversely, the latter poles can be considered
ultraviolet divergences of $\tilde{I}_n$. Some of them lie between the region
of analyticity and the critical dimension $D=4$ which we wish to
analytically continue to.  In the rest of this work we follow a
suggestion of Kaplan, Savage and Wise~\cite{Ka98A,Ka98B} and consider
the effects of cancelling out poles of $\Gamma(\frac{3 - D - 2n}{2})$
(ultraviolet singularities) and $\Gamma(\frac{2n + D - 1}{2})$
(infrared singularities) in the integral $\tilde{I}_n$. This procedure
leads to regularization schemes which differ from the standard
dimensional regularization with minimal subtraction result of
Ref.~\cite{Ka96}, Eq.~(\ref{eq:mins}).  

Of course, after renormalization the on-shell piece of the amplitude
$T$ calculated via Eq.~(\ref{eq:LSE}) is necessarily the same in any
scheme. Otherwise one of the basic tenets of effective field theory,
insensitivity to short-distance physics, is violated.  So we must
ultimately get the same physical result regardless of how many poles
we subtract~\footnote{This point has been stressed in the context of
effective field theories of the nucleon-nucleon interaction by van
Kolck~\cite{vK98}.}. This equivalence after renormalization is
enforced by making appropriate choices for the coefficients $C_{2n}$.
Thus, the behavior of the coefficients, and hence the power-counting
for these coefficients, is affected by the scheme we use to define
$\tilde{I}_n$.  As we will show in the next section some schemes are
more useful for reproducing certain physics in the effective theory.

Consider first the gamma function poles corresponding to ultraviolet
divergences in cutoff regularization.  When $D$ is close to $3 -
2m$, with $m \geq 0$, the poles of $\Gamma(\frac{3 - D - 2n}{2})$
between $D=3-2n$ and $D=4$ have the structure

\begin{equation}
\frac{M}{4 \pi^2} \frac{\mu^{2 m + 1}}{2^{2m}} p^{2(n - m)} 
\frac{1}{D - 3 + 2m}.
\end{equation}
Thus we can cancel these poles out by 
defining, 

\begin{equation}
\tilde{I}_n^{\rm new} \equiv \tilde{I}_n(\mu)
-\frac{M}{2 \pi^2} \sum_{m=0}^n \left(\frac{\mu}{2}\right)^{2m + 1}
p^{2(n-m)}\frac{1}{D - 3 + 2m}
\end{equation}
where $\tilde{I}_n(\mu)$ is given by expression (\ref{eq:In}). 

Similarly, to cancel the poles of $\Gamma(\frac{3 - D - 2n}{2})$
in $D > 4$ dimensions we must add a term 

\begin{equation}
\frac{M}{2 \pi^2} \sum_{m=1}^\infty p^{2(n + m)}
\left(\frac{\mu}{2}\right)^{1-2m} \frac{1}{2m + 3 - D}.
\end{equation}
Thus if we wish to cancel all poles of the expression (\ref{eq:In}) for
dimensions $D > 3 - 2n$ we must define

\begin{equation}
\tilde{I}_n^{\rm{uvPDS}} \equiv \tilde{I}_n(\mu)
- \frac{M}{2 \pi^2} \sum_{m=-\infty}^{n} 
\left(\frac{\mu}{2}\right)^{2m + 1} p^{2(n-m)} \frac{1}{D - 3 + 2m}.
\label{eq:Intildenew}
\end{equation}
This subtracts the poles which correspond to ultraviolet divergences
for $D > 3 - 2n$. We refer to the resulting regularization
scheme as dimensional regularization with ultraviolet power-law
divergence subtraction, or uvPDS.

It is now a straightforward matter to show that uvPDS is precisely
equivalent to cutoff regularization in $D=4$.  For the pieces of the
sum with $m \geq 0$ we find

\begin{equation}
-\frac{M}{2 \pi^2} \sum_{m=0}^n \left(\frac{\mu}{2}\right)^{2m + 1}
p^{2(n-m)}\frac{1}{1 + 2m}=
-\frac{M}{2 \pi^2}
\left[\left(\frac{\mu}{2}\right) p^{2n} + \left(\frac{\mu}{2}\right)^3
\frac{p^{2n-2}}{3} + \ldots +
\left(\frac{\mu}{2}\right)^{2n+1} \frac{1}{2n+1}\right]
\label{eq:mgtzero}
\end{equation}
and for the pieces which arise from poles in $D>4$ it follows from the
observation that

\begin{equation}
\frac{M}{2 \pi^2} \sum_{m=1}^\infty p^{2 m}
\left(\frac{\mu}{2}\right)^{-2m+1} \frac{1}{2m - 1}=
\frac{Mp}{4 \pi^2} \log \left(\frac{\mu/2 + p}{\mu/2 - p}\right),
\label{eq:logsum}
\end{equation}
provided that $p$ is kept less than $\mu/2$.  It is clear that if we
identify $\beta={\mu}/{2}$, then in $D=4$ $\tilde{I}_n^{\rm{new}}$ in
uvPDS is precisely equivalent to the cutoff result of
Eq.~(\ref{eq:integralcutoff}).

\section{Scaling of effective field theory coefficients}

The choice of $\tilde{I}_n$ made by Kaplan {\it et al.}, known as
power-law divergence subtraction, or PDS, is to cancel the pole of
expression (\ref{eq:InKSW}) in $D=3$ for all $n$~\cite{Ka98A,Ka98B}.
Of course, as we have seen the PDS scheme is only one of a much more
general class of definitions of the integral $\tilde{I}_n$.  Kaplan
{\it et al.}'s main motivation in considering the PDS scheme was to
produce a consistent power counting in non-relativistic effective
theories where there is an unnaturally large scattering length.  Much
blood, sweat, and ink has been spilt over this problem during the past
few years~[2--26].
%\cite{Ka98A,Ka98B,Ka96,vK98,Or96,dA97,Ad97,Co97,PC97,Le97,Ka97,LM97,Sc97,Ph97,Ri97,Be97B,vK97,Pa97,Pa98,SF98A,SF98B,Bi98B,Ge98A,Ge98B,CH98A,CH98B,Ep98B,Ep98C,MS98}.
The PDS approach to this difficulty is elegant, and leads to good
power counting for the coefficients in the low-energy Lagrangian.
Here we show that in fact PDS is the simplest scheme of this type
which leads to good power-counting for the bare coefficients $C_{2n}$.
It has been shown how to obtain similar power-counting for the
on-shell T-matrix within the framework of any regularization
scheme~\cite{vK98,vK97}. However, Refs.~\cite{vK98,vK97} do not
address the issue of the scaling of the coefficients in the bare EFT
Lagrangian. The power counting of Refs.~\cite{Ka98A,Ka98B,vK98,vK97}
has also been derived within the framework of the Wilsonian
renormalization group~\cite{Bi98B}.

\subsection{Review of coefficient scaling with the PDS definition of the
divergent integral}

The PDS choice is to retain only the $m=0$ term of the sum
(\ref{eq:Intildenew}). This corresponds to the linear divergence in
the cutoff approach. The integral $\tilde{I}_n$ is defined 
to be (throughout this section we work in $D=4$):

\begin{equation}
\tilde{I}_n^{\rm{PDS}}=-p^{2n} \frac{M}{4 \pi}(ip + \frac{\mu}{\pi}).
\end{equation}
where our result for the second term differs by a factor of $\pi$ from
that of Refs.~\cite{Ka98A,Ka98B} because the angular
integration was performed in $D=4$ rather than $D=3$.

This definition of $\tilde{I}_n$ leads to a straightforward 
solution of the Lippmann-Schwinger equation, (\ref{eq:LSE}):

\begin{equation}
\frac{1}{T^{\rm{on}}(p)}=\frac{1}{\sum_{n=0}^\infty C_{2n} p^{2n}}
+ \frac{M \mu}{4 \pi^2} + \frac{i M p}{4 \pi}.
\label{eq:TPDS}
\end{equation}
Now, we attempt to match this to a form of the inverse amplitude which
corresponds to the presence of an unnaturally large scattering length

\begin{equation}
\frac{1}{T^{\rm{on}}(p)}=-\frac{M}{4 \pi}\left(-\frac{1}{a} 
+ \frac{1}{2} r_e p^2 +O({p^4}) - i p \right),
\label{eq:Tmatching}
\end{equation}
with $1/a \ll \Lambda_0$, and $r_e \sim 1/\Lambda_0$, where $\Lambda_0$ is
the ``natural'' scale set by the theory underlying the low-energy
effective theory. Matching expressions (\ref{eq:TPDS}) and
(\ref{eq:Tmatching}) for $\Lambda_0 > p > 1/a$ it is straightforward
to deduce~\cite{Ka98A,Ka98B} that the coefficients scale as

\begin{equation}
C_0 \sim \frac{1}{M \mu}; \qquad C_{2n} \sim \frac{1}{M \mu^{n+1} \Lambda_0^n},
\label{eq:PDSscaling}
\end{equation}
provided that the scale $\mu$ is kept greater than, or of the order
of, the scale $1/a$. As pointed out by Kaplan {\it et al.} we may now
choose the scale $\mu$ to be of order $p$. 

With this choice all loops involving the operator $C_0$ are of the
same order and so the operator $C_0 (\psi^\dagger \psi)^2$ must be
treated non-perturbatively. This power counting can be understood from
the viewpoint of the renormalization group~\cite{Ka98B,Bi98B} as an
expansion around a nontrivial IR fixed point. This fixed point
describes systems with a bound state at exactly zero energy. In the
PDS scheme, the potential at the fixed point is independent of energy
or momentum and scales like $1/\mu$, as in Eq.~(\ref{eq:PDSscaling}).

{}From Eq.~(\ref{eq:PDSscaling}) we see that the effects of the operators
  $C_{2n} \hat{p}^{2n}$ are suppressed by a factor $(p/\Lambda_0)^n$
  relative to the non-perturbative effects of the dimension-six
  operator. Therefore, these higher-dimensional operators may be
  treated perturbatively.  Hence we can re-expand the first term in
  Eq.~(\ref{eq:TPDS}) in the form
\begin{equation}
\frac{1}{\sum_{n=0}^\infty C_{2n} p^{2n}}=\frac{1}{C_0} - \frac{C_2}{C_0^2} p^2
+ \left[\left(\frac{C_2}{C_0^2}\right)^2 - \frac{C_4}{C_0^2} \right] p^4 
+ \ldots .
\end{equation}
The effective range expansion (\ref{eq:Tmatching}) can then be
reproduced to any desired order in this $p/\Lambda_0$ expansion.

In an RG approach the coefficients in the effective range expansion
are in one-to-one correspondence with the coefficients of the RG
eigenfunctions in the expansion of the potential around the fixed
point. These coefficients scale with definite powers of the
renormalization scale $\mu$ in the PDS scheme or the cut-off in a
Wilsonian approach~\cite{Bi98B}. Although the scaling behavior is the
same in the two approaches, the detailed form of the potential is not:
the fixed-point and RG eigenfunctions all have a more complicated
energy dependence when a cut-off is used.

\subsection{Scaling in PDS in a natural theory}

Note that in a natural theory, where $1/a \sim \Lambda_0$, matching
(\ref{eq:Tmatching}) to (\ref{eq:TPDS}) in fact yields
\begin{equation}
C_{2n} \sim \frac{1}{M \Lambda_0^{2n+1}}
\end{equation}
provided only that $\mu < \Lambda_0$. Since the scale $\mu$ does not
enter these scaling relations we are free to choose $\mu=0$, which is,
of course, the choice of minimal subtraction. However $\mu \neq 0$
with $p \sim \mu$ is also a choice that leads to good power-counting,
Clearly the tree-level effects of the operators $C_{2n}$ are
suppressed by a power $(p/\Lambda_0)^{2n}$. Moreover, the effect of a
loop with superficial degree of divergence $2n + 1$ is down by a
factor of $(p/\Lambda_0)^{2n+1}$, relative to the tree-level $C_0$. Thus
in the natural theory there is no need to treat the effects of $C_0$
non-perturbatively, and the effects of higher-derivative operators
are more strongly suppressed than in the unnatural case.

\subsection{Scaling in uvPDS}

\begin{figure}[h,t,b]
   \vspace{0.5cm}
   \epsfysize=2.5cm
   \centerline{\epsffile{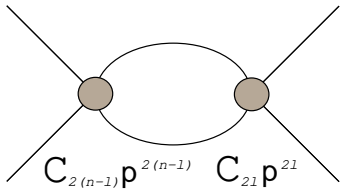}}
   \centerline{\parbox{11cm}{\caption{\label{fig3}
Characteristic loop graph.
  }}}
\end{figure}

We can now deduce the effect of modifying the definition of
$\tilde{I}_n$ by adding the terms for $m=1$ to $m=n$ which appear in
Eq.~(\ref{eq:Intildenew}). Recall that if a cutoff regulator is used
these terms correspond to cubic and higher power-law divergences.
Attempting to solve the theory exactly, as was done above for the PDS
subtraction, becomes much more complicated (see
Refs.~\cite{Ph97,Ri97,Ge98A} for examples), as each integral
$\tilde{I}_n$ has a different number of subtractions.
This complexity gives rise to nonlinear relations between the bare and
renormalized coefficients.  However, the crucial point in any such
attempt is that if we keep the scale $\mu$ well below the scale
$\Lambda_0$ then the effect of all power-law divergences will be
suppressed by powers of $\mu/\Lambda_0$, and so they make no change to
the scaling of coefficients quoted above. For instance, if we examine
a loop as shown in Figure~\ref{fig3} with one insertion of
$C_{2(n-l)}$ and one of $C_{2l}$ then, considering for the present
only the pieces of the sum in Eq.~(\ref{eq:Intildenew}) corresponding
to power-law divergences of degree three and above, the loop becomes

\begin{equation}
C_{2(n-l)} C_{2l} \frac{M}{2 \pi^2} \sum_{m=1}^{n} 
\left(\frac{\mu}{2}\right)^{2m + 1} p^{2 (n-m)} \frac{1}{1 + 2m}.
\label{eq:sumform}
\end{equation}
Given the scaling above for the coefficients $C_{2j}$ the $m$th term
here produces scaling of the coefficient $C_{2(n-m)}$ of the form

\begin{equation}
C_{2(n-m)} \sim \frac{1}{M \mu^{n-2m+1} \Lambda_0^n}.
\label{eq:loopscaling}
\end{equation}
Compared to the leading scaling of this coefficient $C_{2(n-m)} \sim
1/(M \mu^{n-m+1} \Lambda_0^{n-m})$ the scaling (\ref{eq:loopscaling})
is suppressed by a factor of $(\mu/\Lambda_0)^m$. Thus, even if the
terms in Eq.~(\ref{eq:Intildenew}) which correspond to power-law
divergences of higher degree than those considered in PDS were
included they would not modify the PDS scaling (\ref{eq:PDSscaling}).

By contrast, the terms arising from poles in $D=5$ and above, which
sum up to produce the logarithm of Eq.~(\ref{eq:logsum}), {\it would}
modify the PDS scaling (\ref{eq:PDSscaling}) if they were included in
the definition of $\tilde{I}_n$. To see this let us only retain the
terms of the sum in Eq.~(\ref{eq:Intildenew}) for $m \leq 0$. Then

\begin{equation}
\tilde{I}_n=-p^{2n} \frac{M}{4 \pi} \left(ip + \frac{\mu}{\pi} - \frac{1}{\pi}
\log\left(\frac{\mu + 2 p}{\mu - 2 p}\right)\right),
\end{equation}
so solving the Lippmann-Schwinger equation gives

\begin{equation}
\frac{1}{T^{\rm{on}}(p)}=\frac{1}{\sum_{n=0}^\infty C_{2n} p^{2n}}
+ \frac{M \mu}{4 \pi^2} 
- \frac{M p}{4 \pi^2} \log \left(\frac{\mu + 2 p}{\mu - 2 p}\right)
+ \frac{i M p}{4 \pi}.
\label{eq:TCODS}
\end{equation}
Expanding out the log in powers of $p/\mu$ it is straightforward to
see that matching to the expression (\ref{eq:Tmatching}) requires

\begin{equation}
C_{2n}=\frac{1}{M \mu^{2n+1}}.
\end{equation}
This, of course, is the same scaling one would expect in a cutoff
theory in which the cutoff was kept below the scale
$\Lambda_0$~\cite{Le97,Be97B}. 

Now, if the coefficients do scale in this way then for momenta $p \sim
\mu$ the effect of the higher-derivative operators $C_{2n} p^{2n}$ is
{\it not} suppressed relative to that of the ``lowest-order'' operator
$C_0$. Thus if uvPDS is used to regulate the divergent loop integral
the resulting power-counting for the $C$'s is such that one cannot
justify any truncation of the sum over all higher-derivative
operators. Consequently, we conclude that the choice
(\ref{eq:Intildenew}) does not lead to good power-counting for the
coefficients $C_{2n}$.

The physical reason this occurs is that the inclusion of terms that
contain negative powers of $\mu$ in the definition of the integral
$\tilde{I}_n^{\rm{uvPDS}}$ introduces energy-dependence in the
amplitude at scale $\mu$. Since $\mu$ is much less than the natural
scale of energy dependence $\Lambda_0$, this energy dependence is
unnaturally rapid. It must be cancelled by using larger values of the
coefficients $C_{2n}$. In fact as shown in~\cite{Bi98B}, this occurs
automatically within an RG treatment where the unnatural energy
dependence is absorbed into the forms of the fixed-point potential and
the RG eigenfunctions for the expansion around it. In our case we
would have the fixed-point potential

\begin{equation}
V(p)=\left(-\frac{M \mu}{4 \pi^2} + 
\frac{M p}{4 \pi^2}  \log \left(\frac{\mu + 2 p}{\mu - 2 p}\right)\right)^{-1}.
\end{equation}
If this potential is iterated via the Lippmann-Schwinger equation and
uvPDS (which we have already shown is equivalent to a cutoff) applied
to define the divergent integral then all the spurious
energy-dependence introduced by the terms in the sum for $\tilde{I}_n$
with negative-powers of $\mu$ are cancelled. The analysis given here
shows that this is essentially equivalent to just ignoring the effects of
the poles in $D > 5$ which lead to logarithmic behavior in the inverse
amplitude altogether.

One difference between uvPDS (or a cut-off) and PDS \`a la KSW is that
it distinguishes between energy and momentum dependence in the
potential. The RG eigenfunctions that correspond to the terms in the
effective range expansion are purely energy-dependent~\cite{Bi98B}.
Ultimately they can be thought of a terms in an energy-dependent
pseudopotential that acts as an energy-dependent boundary condition on
the wave function at the origin~\cite{vK98,Bi98B}. There are in addition
momentum-dependent eigenfunctions with different scaling behaviors,
but these do not contribute to the on-shell scattering amplitude.

\section{Summary}

The integral $\tilde{I}_n$ is a key ingredient of non-relativistic
effective field theories in general, and EFTs of the nucleon-nucleon
interaction in particular. Since this integral has infinitely many
poles in the complex space of dimensions there are infinitely many
ways to define it, depending on how many of these poles we choose to
subtract in our redefined $\tilde{I}_n$. The original suggestion of
Ref.~\cite{Ka98A} which removes the pole at $D=3$ therefore exists
amidst myriad alternative schemes. The alternative corresponding to
removing {\it all} of the poles which arise due to ultraviolet
divergences of $\tilde{I}_n$ exactly reproduces the result obtained by
simply regulating $I_n$ via a cutoff.

We have also analyzed the way that coefficients in the effective field
theory Lagrangian scale in regularization schemes using these more
general definitions of $\tilde{I}_n$. This analysis shows
that, in the presence of an unnaturally large scattering length, PDS
is the simplest definition of a subtracted $\tilde{I}_n$ which leads
to good power-counting for the coefficients that appear in the
effective field theory Lagrangian.

\section*{Acknowledgments}

D.~R.~P. acknowledges useful discussions on the nature of dimensional
regularization with I.~R.~Afnan, T.~D.~Cohen, and A.~Schreiber. D.~R.~P. and
S.~R.~B. thank the U.~S. Department of Energy, Nuclear Physics
Division, for its support (grant DE-FG02-93ER-40762). M.~C.~B. 
is grateful for useful comments from J.~A.~McGovern, and for 
support from the U.~K. EPSRC.

\appendix

\section{``Maximal'' subtraction}

Here we consider the role played by the poles in $D < 1 - 2n$. As
mentioned above, these correspond to infrared singularities of the
original integrand.  Since the integral of interest has no infrared
divergences in $D=4$, this discussion is not particularly relevant to
nature. However, it leads to an intriguing result, and so we present
it in this appendix.

To cancel the poles of $\Gamma(\frac{2n + D - 1}{2})$ in $D < 1-2n$
dimensions we must add a term

\begin{equation}
-\frac{M}{2 \pi^2} \sum_{m=n+1}^{\infty} 
\left(\frac{\mu}{2}\right)^{2m + 1} p^{2(n-m)} \frac{1}{D - 3 + 2m}.
\end{equation}
We can then define a scheme which we call maximal subtraction,
that subtracts {\it all} poles in the complex $D$-plane:

\begin{equation}
\tilde{I}_n^{\rm{all}} \equiv \tilde{I}_n(\mu)
- \frac{M}{2 \pi^2} \sum_{m=-\infty}^{\infty} 
\left(\frac{\mu}{2}\right)^{2m + 1} p^{2(n-m)} \frac{1}{D - 3 + 2m},
\label{eq:Intildenewall}
\end{equation}
in all dimensions $D$ in the complex $D$-plane. We now specialize to 
the case $D=4$.

The sum in Eq.~(\ref{eq:Intildenewall}) is divergent for any value of the
momentum $p$. In order to define a sum for this series we first define
the function $f(p)$ via

\begin{equation}
f(p)=- \frac{M}{2 \pi^2} \sum_{m=0}^{\infty}
\left(\frac{\mu}{2}\right)^{2m + 1} p^{2(n-m)} \frac{1}{2m+1}.
\end{equation}
If $\mu/2 < p$ then this series may be summed to yield 

\begin{equation}
f(p)=-\frac{Mp}{4 \pi^2} {p^{2n}}\log \left(\frac{p+\mu/2}{p-\mu/2}\right). 
\end{equation}
In the region of the complex $p$-plane, $\mu/2 < |p| < R$, with $R$
some large real number, this function is an analytic function of
$p$. Hence we can define a function $\tilde{f}(p)$ which is the
analytic continuation of $f(p)$ into the region $0 \leq |p| < R$. We make
the analytic continuation

\begin{equation}
\tilde{f}(p)=-\frac{Mp}{4 \pi^2} p^{2n} \log\left(\left|
\frac{p+\mu/2}{p-\mu/2}\right|\right) + p^{2n} \frac{i M p}{4 \pi}.
\end{equation}

We now define $\tilde{I}_n^{\rm{MaxS}}$ to be the result found if the
sum over positive $m$ in Eq.~(\ref{eq:Intildenewall}) is replaced by its
analytic continuation $\tilde{f}$, thus:

\begin{equation}
\tilde{I}_n^{\rm{MaxS}}=\tilde{I}_n(\mu) +  p^{2n} \frac{i M p}{4 \pi}.
\end{equation}
Taking the limit $D \rightarrow 4$ and evaluating the (finite)
integral $\tilde{I}_n(\mu)$ implies

\begin{equation}
\tilde{I}_n^{\rm{MaxS}}=0.
\end{equation}
By subtracting all the poles in the complex $D$-plane we have forced a
definition of the value of the divergent integral $I_n$ which gives
zero.

\end{document}